\documentstyle[aps,epsf]{revtex}
\def\`#1{\if#1i{\accent18 \i}\else{\accent18 #1}\fi}
\def\'#1{\if#1i{\accent19 \i}\else{\accent19 #1}\fi}

\begin{document}
\draft
\date{\today}
\title{Air Showers and Geomagnetic Field.
}
\author{A.~Cillis and S.~J.~Sciutto}
\address{Laboratorio de F\'isica Te\'orica\\
Departamento de F\'isica\\
Universidad Nacional de La Plata\\
C. C. 67 - 1900 La Plata\\
Argentina}

\maketitle

\begin{abstract}
The influence of the geomagnetic field on the development of air
showers is studied.
The well known International Geomagnetic Reference Field was included
in the AIRES air shower simulation program as an auxiliary tool to allow
calculating very accurate estimations of the geomagnetic field given the
geographic coordinates, altitude above sea level and date of a given
event.
Our simulations indicate that the geomagnetic deflections alter
significantly some shower observables like, for example, the lateral
distribution of muons in the case of events with large zenith angles
(larger than 75 degrees).
On the other hand, such alterations seem not to be important for smaller
zenith angles. Global observables like total numbers of particles or
longitudinal development parameters do not present appreciable dependences
on the geomagnetic deflections for all the cases that where studied.
\end{abstract}

\pacs{96.40.Pq, 13.10.+q, 02.70.Lq}

\section{Introduction}

The understanding of the origin and nature of high energy cosmic rays is
one of the most challenging topics in contemporary astrophysics. With the
knowledge currently available we cannot discard the possibility that the
highest energy primary particles (those with energies above $10^{19}$ eV)
have an entirely different origin than lower energy cosmic rays. This
generates a variety of questions that cannot be definitively solved
without adequate sets of experimental data. For this reason, several
projects have been envisioned to determine the main characteristics of
such cosmic rays.

The AUGER Observatory is one of these projects \cite{Auger}. It consists
in two identical hybrid detectors, respectively located in the 
Northern and Southern Hemisphere to get full sky coverage. Each hybrid
detector consists in a surface array and a fluorescence detector,
optimized to measure different parameters of the particle air showers
generated after the incidence of high energy cosmic rays into the Earth's
atmosphere. The Observatory will be able to measure the arrival
direction of the primary particle and the electromagnetic and muonic
components of the showers at ground level. The longitudinal profile of the
shower (number of charged particles as a function of the altitude) will be
also available for approximately 10 \% of the measured events. These
special events are referred as ``hybrid events''.

To clearly understand the relationship between the characteristics of the
primary particle (energy, mass, etc.) and the quantities that will be
measured by the AUGER detectors, it is essential to study the shower
development by means of computer Monte Carlo simulations. We started
working in this subject some years ago, and we have developed a set of
programs to simulate air showers and manage all associated output data.
Such simulating system is identified with the name AIRES (AIRshower
Extended Simulations) \cite{Aires,HM,PRD}.

The first version of AIRES was developed on the basis of the well known
MOCCA program, created by A. M. Hillas for the Haverah Park experiment
\cite{mocca}. The AIRES program incorporates substantial improvements,
in particular from the computational point of view. Later versions do also
include additional physics algorithms that complement the original set of
procedures taken from MOCCA. 

The particles currently processed by AIRES are: gamma rays, electrons and
positrons, muons, pions, kaons, nucleons, anti-nucleons and nuclei.
Neutrinos are generated  (in pion or muon decays, for example) and
accounted for their energy, but not tracked.
The shower particles can undergo the following processes: $(1)$ Electromagnetic interactions: Pair production, bremsstrahlung
(electrons and positrons), Compton and photoelectric effects and emission
of knock-on electrons ($\delta$ rays). The LPM effect and dielectric
suppression that affect high energy production and bremsstrahlung
processes are taken into account using recently developed procedures (for
details see reference \cite{PRD}). $(2)$ Hadronic interactions: Nuclear fragmentation and inelastic
collisions, managed via calls to external packages like SIBYLL
\cite{sibyll} or QGSJET \cite{qgsjet}. $(3)$ Unstable particle decays. 
$(4)$ Particle propagation includes continuous energy losses (ionization)
and scattering. The curvature of the Earth is always taken into account
(it is possible to process showers with zenith angles in the full range
[$0^{\circ}$, $90^{\circ}$]), as well as geomagnetic deflections.

The purpose of this work is to analyze the effect of the Geomagnetic Field
(GF) on the development of the showers, giving quantitative estimates of
the influence of the GF on the very high energy showers using environmental
conditions similar to the ones of the AUGER observatory.

We have first studied the principal characteristics of the GF: Its origin,
magnitude and variations. Thereafter we have analyzed the different GF
models that are normally used: The dipolar models \cite{campoB} which consider 
the GF as generated by magnetic dipoles; and the
so-called International Geomagnetic Reference Field (IGRF) \cite{IGRF}, a
more elaborated model, based on a high-order harmonic expansion whose
coefficients are fitted with data coming from a network of geomagnetic
observatories all around the world.

One of the conclusions that comes out from our analysis is that the
dipolar models are not useful to evaluate the GF at a given arbitrary
location with enough accuracy. On the other hand, the predictions of
the IGRF proved to match with the corresponding experimental data with
errors that are at most a few percent. For example, the difference
between the IGRF predictions on the field components and the
experimental data are less than 500 nT (nanotesla, 1 ${\rm nT}=10^4$
gauss).  We have therefore selected the IGRF to link it to the
simulation program AIRES, as an adequate model to synthesize the GF at
a given geographical location and time.

This paper is organized as follows: In section 2 we report on the
simulations performed to analyze the influence of the GF on the shower
development and in section 3 we place our final remarks and conclusions.
In appendix A we give a short description of the GF and we briefly
comment different models that exist at present and discuss the practical  
implementation of the GF in the AIRES program.
Additional information needed to analyze the influence of the GF without 
other distortions, like the ones introduced by the geometry (the axis of 
the showers are usually inclined) and the attenuation in the atmosphere, 
is given in appendix B.

\section{Simulations}

We have analyzed the influence of the GF on air shower observables performing
some simulations in a variety of initial conditions.

We have simulated several thousands of showers with varying primary energy
(in the range of ultra-high energies) and zenith angle.
We have considered different sites in our simulations, which imply
different intensities and directions of the GF. In most cases these
environmental conditions correspond to the Northern and Southern 
sites for the AUGER observatory, which are, respectively Millard County
(Utah, USA) and El Nihuil (Mendoza, Argentina) \cite{Auger}.

It is interesting to note that the average GF intensity at the
Millard site is twice the corresponding one for the El Nihuil Southern
site (50000 nT and 25000 nT, respectively). As a consequence, the
influence of the field on the particles' paths will be different at each
site, being more appreciable at Millard where the GF is more intense.

Our analysis is mainly based on comparisons of average values of
observables coming from simulations performed with the program AIRES
alternatively taking and not taking into account the GF deflections of
charged particles (For a  more detailed description of the implementation
of the GF in AIRES, see appendix A).

The direct inspection of global observables such as the longitudinal
development of all charged particles (total number of charged particles
plotted against the atmospheric depth, $X$), shower maximum (atmospheric
depth where the total number of charged particles is maximum), total
number of particles at ground, etc, ~shows no evident effect of the GF.

To illustrate this point we have plotted in figure \ref
{fig:longigp81} the longitudinal development of the number of all charged
particles and muons for showers initiated by  $10^{20}$ eV gammas (a) and
protons (b).
The zenith angle is $81.5^{\circ}$ ($\cos \Theta=0.15$), the azimuth is
fixed at $90^{\circ}$ to ensure a strong effect of the GF, and the
environmental conditions are those of Millard site.
Comparing the plots coming from the simulations with (solid line) and
without (dashed line) GF, it is evident that there are no significant
differences due to the GF deflections for both charged particles and muons
cases. One of the main implications that can be derived from these figure
is that the GF does not induce any appreciable variations in both the
position of the maximum ($X_{max}$) and the maximum number of charged
particles ($N_{max}$). 
We recall that these observables are the main quantities that can be
measured with the fluorescence detector, and are essential for determining
the shower energy and primary composition.

Although the differences between the simulations taking  and not taking
into account the GF are not significant for global observables, as we have
shown before, some variations do appear when a detailed analysis of some
particle distributions is made. This is the case, for example, of the
density distribution of muons at ground level (lateral distribution of
muons).

This kind of particles is the most affected by the GF. This is
due to the fact that muons usually travel long distances without
interacting with the medium, allowing for bigger GF deflection angles.
Additionally, for large zenith angles, the muonic component of the shower
generally represents an important fraction of the measurable ground level
signal. Therefore this case should be studied in detail in order to
establish whether or not the standard ground measurement techniques
\cite{Auger} (developed for nearly vertical showers) are valid when the
zenith angle not small.

The plots in figure \ref {fig:longigp81} illustrate how the muonic
component of the showers becomes progressively more important (in
comparison with all the charged particles) as long as the shower continues its
evolution after having reached its maximum.

The ratio between the numbers of muons and electromagnetic particles, that
is, 
\begin{equation}
\eta=\frac{\hbox{Number of } \mu ^{+/-}}{\hbox{Number of } \gamma+
\hbox{Number of } e^{+/-}}
\label{eta}
\end{equation}
gives a convenient quantitative measure of the relative importance of the
muonic component in a determined case.

In figure \ref {fig:ratiovszenithe3}, the ratio $\eta$, evaluated at
ground level, is plotted versus the zenith angle for the typical case of
$3\times10^{20}$ eV proton showers. The ground level altitude is 1400
m.a.s.l.~ $\eta$ is practically constant for zenith angles lower than 60
degrees. Around this point $\eta$ begins to rise abruptly (the parameter
grows more than 2 orders of magnitude when the zenith angle passes from
50 to 70 degrees)  and reaches a maximum at approximately 75 degrees.
$\eta(70^{\circ})$ is some 400 times larger than $\eta(0^{\circ})$. Beyond
that point $\eta$ decreases slightly.

There is another remarkable experimental reason to study the influence of
the GF on the lateral distribution of muons for large zenith angles: The
total average signal observed in the \u Cerenkov detectors 1.2 m
depth (as in the  case of the Auger Observatory \cite{Auger}) is
dominated by the muonic component which is twice the signal of the
electromagnetic particles for zenith angles of 80 degrees, that is, a
fraction tree times larger than the corresponding one for a zenith of  
$30^{\circ}$.

In the remaining part of this section we are going to present some
representative results obtained from the simulations performed with the
AIRES program. We have analyzed a variety of initial conditions: Two
geographical locations (El Nihuil and Millard sites) each one with
different GF, zenith angles in a wide range, from $0^{\circ}$ to
$81.5^{\circ}$, and ultra-high energies in the range $10^{19}$ eV to
$10^{20.5}$ eV.
For very inclined showers the effect of the GF  deflections on the lateral
distribution of muons is very significant. This can be put into evidence 
comparing figures \ref {fig:g100z81conb_mu012} and  \ref
{fig:g100z81sinb_mu012}.
These figures corresponds to $10^{20}$ eV gamma showers and the injection
altitude is the top of the atmosphere. In these plots, the muon density
is represented at each case by means of diagrams and contour line plots.
The plots labelled (a) represent the row ground densities, while the plots
labelled (b) and (c) represents the geometrical (equation (\ref {rho_0}))
and geometrical with attenuation correction (equation (\ref {rho_g}))
projections onto the shower front plane, respectively. The plots of figure
\ref {fig:g100z81sinb_mu012} (GF turned off) show that the well known
distortion that makes contour lines approximately elliptical can be
eliminated after applying the procedures of appendix B: The contour lines
of figure \ref {fig:g100z81sinb_mu012} c are approximately circles, 
concentric with the shower axis. 
The contour lines of figure \ref {fig:g100z81sinb_mu012} b, are also
approximately circular, but a careful analysis shows that they are
slightly eccentric (in this figure the centers are shifted towards the
right).
This typical eccentricity indicates that the attenuation correction
(equation (\ref {rho_g})) cannot be neglected.

When the GF is taken into account (figure \ref {fig:g100z81conb_mu012}),
the two dimensional density pattern is different, and the application of
the procedures to project onto the shower front plane permits putting in
clear evidence the lack of cylindrical symmetry of the shower in this
case.

Ground and shower front plane (with all corrections) $\mu^{+}$ and
$\mu^{-}$ distributions are presented in figures \ref
{fig:p100z81conb_mpm02} and \ref {fig:p100z81sinb_mpm02}, in the same
conditions as in figure \ref {fig:g100z81conb_mu012} and
\ref {fig:g100z81sinb_mu012}, but for proton showers and injection altitude
of 200 g/cm$^2$, respectively taking and not taking into
account the GF. In this case the injection altitude is 200 g/cm$^2$ in order
to have enough signal at ground level. In the case of gamma showers, we
have chosen the injection altitude at the top of the atmosphere because this
kind of particles are very penetrating ones due to the LPM effect \cite{PRD}.
The direction of the horizontal field (H) and the projection of the GF onto
the shower front plane are indicated for convenience. When the GF is enabled
(figure \ref {fig:p100z81conb_mpm02}) the separation between $\mu^{+}$ and
$\mu^{-}$ becomes evident. 

The simulations corresponding to the distributions of figures \ref
{fig:g100z81conb_mu012} and \ref {fig:p100z81conb_mpm02} were performed
for very inclined showers (zenith angle $81.5^{\circ}$), and a relatively
intense GF (${\rm F}=52800$ nT).
In the case of El Nihuil site, where the GF intensity is about one half of
the previous one, the effect of the GF deflections is smaller, but not
negligible. 

The distortions generated by the GF for very inclined showers
diminish dramatically as long as the zenith angle is reduced. To
illustrate this point, let us consider the plots of
figure \ref {fig:p60n8em_mpm02}, which corresponds to $3\times10^{19}$ eV
proton showers with zenith angle $60^{\circ}$.
A careful inspection of these graphs permits detecting some little
differences between the $\mu^{+}$ and $\mu^{-}$ distributions.
Such differences, however, compensate noticeably when evaluating the
total ($\mu^{+}$ and $\mu^{-}$) distribution (not plotted here), which is
practically equivalent to the corresponding one coming from the
simulations without GF.

\section{Conclusions}

We have discussed in this work the influence of the geomagnetic field on the
most common observables that characterize the air showers initiated by
astroparticles. The data used in our analysis were obtained from computer
simulations performed with the AIRES program.

Our work includes the analysis of the main properties of the geomagnetic
field, as well as the implementation of the related algorithms in the
program AIRES.

By means of the International Geomagnetic Reference Field (IGRF) it is
possible to make accurate evaluations of the average geomagnetic field at
a certain place given its geographical coordinates, altitude above sea
level and time. We have used this tool to run the simulations using a
realistic geomagnetic field.

The changes that global observables like the longitudinal
development of all charged particles experiment when the
geomagnetic field is taken into account, are generally small.
On the other hand, we have found that in some distributions, like in the
lateral distribution of muons, for the case of realistic fields and for
large zenith angles (larger than 70 degrees), the differences between the
cases where such field is taken or not taken into account become 
significant.

For showers with zenith angles less than 70 degrees, the deflections due
to the GF do not generate important alterations in such distributions,
allowing for safe application of analysis techniques that do not take care
of the effect of the GF deflections.

It is worthwhile to mention that this work is restricted to the study
of the consequences derived from the deflections of charged particles that
move under the influence of the Earth's magnetic field. Other effects
modifying the behavior of air showers and related to the GF will be
considered in future works.

\section*{Acknowledgments}

We are indebted to L. N. Epele, C. A. Garc\'{\i}a Canal, and H. Fanchiotti for
useful discussions; also to O. Medina Tanco (S\~{a}o Paulo University,
Brazil) and J. Valdez (UNAM, Mexico) for their help to obtain information
about the IGRF.

The experimental data from Las Acacias and Trelew Observatories are courtesy
of J. Gianibelli (FCAGLP, La Plata, Argentina).

Finally we want to thank C. Hojvat (Fermilab, USA) who gave us the  
possibility of running our simulations on very powerful machines.

\appendix

\section{Geomagnetic deflections of charged particles}

\subsection{Calculating the Earth's magnetic field}

The Earth's magnetic field is described by seven parameters \cite{campoB},
namely, total intensity (F), inclination (I), declination (D), horizontal
intensity (H), vertical intensity (Z), and the north (X) and east (Y)
components of the horizontal intensity. D is the angle between the
horizontal component of the magnetic field and the direction of the
geographical north and I is the angle between the horizontal plane and the
total magnetic field. It is considered positive when the magnetic field
points downwards. Also Z is positive when I is positive \cite{campoB}.

The GF is generated by internal and external sources. The first ones
are related to processes in the interior of the Earth's core and the
intensity of the field generated goes from 20000 to 70000 nT while the
second ones would be related to ionized currents in
the high atmosphere and it contribution is around 100 nT \cite{campoB}. 

The different components of the GF (external and internal) are not
uniform over position and time. The secular and periodic variations
\cite{campoB} are originated by the external field. The firsts go from 10
up to 150 nT per year and the second ones are less than 100 nT.
Also, there are sudden disturbances in the GF (namely, magnetic storms)
which may last from hours up to several days and rarely modify the field
in more than 500 nT.

The simplest way to model the GF is to assume that it is generated by a
magnetic dipole (dipolar central and  eccentric models) \cite{campoB}.
However, when a more accurate reproduction of the field is needed, it is
necessary to go beyond the dipole approximation and make a higher order
harmonic analysis of the GF \cite{campoB}.
Due to the spherical symmetry of the problem, the solution can be
conveniently expressed in terms of the following expansion
\begin{equation}
\phi =a\sum_{n=1}^N \sum_{m=0}^n \left(\frac{a}{r}\right)^{n+1}
[g_{nm}\cos m\lambda +h_{nm}\sin m\lambda] P_n^m(\cos \varphi )
\label{soluc}
\end{equation}
where $a$ is the mean radius of the Earth (6371.2 km), $r$ is the radial
distance from the center of the Earth, $\lambda $ is the longitude
eastwards from Greenwich, $\varphi $ is the geocentric colatitude, and
$P_n^m(\cos \varphi )$ is the associated Legendre function of degree $n$
and order $m$, normalized according to the convention of Schmidt. $N$ is
the maximum spherical harmonic degree of the expansion.
The International Geomagnetic Reference Field (IGRF) \cite{IGRF} is a
series expansion like (\ref {soluc}) where the coefficients  are adjusted  
to fit experimental measurements coming from a network of geomagnetic
observatories located all around the world. Set of spherical harmonic
coefficients ($g_{nm}$ and $h_{nm}, N=10)$ at 5-year intervals starting
from 1900 are currently available. Coefficients for dates between
5-year epochs are obtained by linear interpolation between the corresponding
coefficients for the neighboring intervals.
The error of the field components and the D and I angles are
less than 500 nT and 30 arc minutes, respectively. These errors are
relatively small and this makes the IGRF model a very useful tool to
estimate the GF at any geographic location and any time belonging to its
validity interval.

We have evaluated the results coming from the different models already
mentioned in a variety of situations, in order to establish which of them
is the most convenient to cover the needs arising in an air shower
simulation algorithm. Our conclusion is that the IGRF is the most
convenient model that can give accurate estimations of the GF to be used in
air-shower simulations \cite{paperGF0}. An illustrative example is placed  
in figure \ref {fig:acacia}, where a comparison between experimental data
\cite{acacias} and the IGRF model is plotted versus time. The F, H and
Z components are shown. In all cases, the absolute difference between
estimated and measured fields is always less than 200 nT. The errors in
the estimation of I and D (not displayed) are always less than 0.5
degrees.
More illustrative examples of the different models of the GF are plotted 
in reference \cite{paperGF0}.

\subsection{Practical Implementation}

It is assumed that the shower develops under the influence of a constant and
homogeneous magnetic field which is evaluated before starting the
simulations \footnote{Since the region where the shower develops is very
small when compared with the Earth's volume, the mentioned approximation
of a constant and homogeneous field is amply justified.}.
In order to calculate the GF,  special subroutines using the IGRF model
have been incorporated to the AIRES program \cite{Aires}.

If a charged particle $q$ advances a distance $\Delta s$ in a uniform,
static magnetic field $\mathbf{B}$ ($\Delta s=\beta c\Delta t$, $\beta
=v/c$), the motion can be approximately calculated via \footnote{We use
MKS units.}:

\begin{equation}
\widehat{\mathbf{u}}(t+\Delta t)\cong \widehat{\mathbf{u}}(t)+
\frac{d\widehat{\mathbf{u}}}{dt}%
\Delta t=\widehat{\mathbf{u}}(t)+\left(\frac{qc^2\Delta t}{E}\right)
\widehat{\mathbf{u}}%
\times \mathbf{B}   \label{u}
\end{equation}

where $\widehat{\mathbf{u}}=\frac{\mathbf{v}}
{\mathbf{\left|v\right| }}$ is the unit velocity vector, $\mathbf{v}$ is
the
velocity of the particle, $E$ is the total energy of the particle (rest
plus kinetic) and $c$ is the speed of light.

For this approximation to be valid, it is needed that

\begin{equation}
\omega \Delta t=\frac{\omega \Delta s}\beta \ll 1  \label{w2} 
\end{equation}
where
\begin{equation}
\omega =\frac{qc^2B}E  \label{w}
\end{equation}
is the angular velocity of the particle \cite{jackson}. 

The magnetic deflection algorithm implemented in AIRES makes use of equation
(\ref {u}) (taking care that (\ref {w2}) is always satisfied) to
evaluate the updated direction of motion at time $t+\Delta t$.
However, it also uses a ``technical trick'', inspired in a similar procedure
used in the well-known program MOCCA \cite{mocca}: The path $\Delta s$ is
divided in two halves of length $\Delta s/2$ each. Then the particle is
moved the first half using the old direction of motion
$\widehat{\mathbf{u}}(t)$, and
the second one with the updated vector $\widehat{\mathbf{u}}(t+\Delta
t).$ 

\section{Geometry and Attenuation in the atmosphere}

To analyze adequately the modifications introduced by the GF deflections
in the shower observables, it is convenient to subtract the deviation
introduced geometrically (due to the fact that the shower axes are usually
inclined) and related to the atmospheric attenuation.
In figure (\ref {fig:geometry}), the basic geometrical elements
corresponding to an inclined shower are represented schematically. Let 
$\Theta$ and $\Phi$ be the shower zenith and azimuth angles, respectively.
If $(r,\varphi)$ are the polar coordinates of P in the ground
plane and $(r_0,\beta)$ represent the polar coordinates of the same point
with respect to the shower front plane (plane perpendicular to the shower
axis containing P), it is easy to show that
\begin{equation}
r_0 = \overline{{\bf PQ}} = r\,
\sqrt{1-\sin^2\Theta\cos^2(\varphi-\Phi)}
\label{ro}
\end{equation}
and
\begin{equation}
\tan\beta=\frac{\tan(\varphi-\Phi)}{\cos\Theta}
\label{tan}
\end{equation}

In general, the density of a given type of particles (for example muons)
at ground level will be a function of $r$ and $\varphi$,
$\rho_{g}(r,\varphi)$.
On the oder hand, showers that develop with no GF deflections will have
cylindrical geometry with respect to the shower axis and the density
measured on the shower front plane will not depend on $\beta$:
$\rho_{0}(r_{0},\beta)$=$\rho_{0}(r_{0})$.
For vertical showers, $\rho_{0}$ and $\rho_{g}$ refer to the same
magnitude, but when the zenith angle is not zero, $\rho_{0}$ is
different from $\rho_{g}$.

If the variation on the total number of particles with the atmospheric
depth is neglected, then $\rho_{0}$ and $\rho_{g}$ are related by a simple
geometrical transformation:
\begin{equation}
\rho_{0}(r_0,\beta) =\frac{\rho_g(r,\varphi)}{\cos\Theta} 
\label{rho_0} 
\end{equation}
(in this formula, $\rho_{0}$ will not depend on $\beta$ for cylindrically
symmetrical showers).
It is well known, however, that the number of shower particles is not
constant for different depths. This means that the density $\rho_0$
depends on $X$, the depth of point {\bf Q}, and therefore equation (\ref
{rho_0}) will not be approximate enough in these cases.

A transformation between $\rho_0$ and $\rho_g$ that proves to work
acceptably well in most practical cases can be easily derived accepting
the following empirical parameterization for the lateral distribution at
varying atmospheric depth \cite{Auger}
\begin{equation}
\rho(r_0,X) = k(X)\, r_0^{\eta(X)}\, f_0(r_0)
\label{rho}
\end{equation}
where $k$ and $\eta$ are functions of $X$, and $f_0$ only depends on
$r_0$.

Let $X_g$ be the vertical depth of the ground plane. The
vertical depth at point {\bf Q}, $X$ is usually not much different
from $X_g$, and can be excellently approximated by means of a
locally isothermic atmosphere:
\begin{equation}
X \simeq X_g e^{-\xi z_Q}
\label{X}
\end{equation}
where $\xi$ is a parameter characterizing the atmosphere which can be
determined easily, and 
\begin{equation}
\quad z_Q =
\frac{r}{2}\sin(2\Theta)\cos(\varphi-\Phi)
\label{zQ}
\end{equation}
is the altitude of point {\bf Q} (with respect to the ground level)

From equation (\ref {rho}) it is easy to show that
\begin{equation}
\frac{\rho_0(r_0,X)}{\rho_0(r_0,X_g)} = \frac{k(X)}{k(X_g)}\,
r_0^{[\eta(X)-\eta(X_g)]}
\label{rho/rho}
\end{equation}

When $X$ $>$ $X_{\rm max}$ the shower is in its attenuation phase, and
$k(X)$ (related to the total number of particles at depth $X$) can be
approximated as ($X\simeq X_g$):
\begin{equation}
\frac{k(X)}{k(X_g)} \simeq e^{-a(X-X_g)}
\label{k/k}
\end{equation}
In the same conditions, $\eta(X)$ can be satisfactorily represented as a
linear function of $X$
\begin{equation}
\eta(X)-\eta(X_g)\simeq b(X-X_g)
\label{eta2}
\end{equation}
$a$ and $b$ are parameters to be determined. Notice that $a$ and $b$
do not depend on $r_0$.
Equations (\ref {rho/rho}), (\ref {k/k}) and (\ref {eta2}) yield
\begin{equation}
\rho_0(r_0,X)=\rho_0(r_0,X_g) \exp\left[-(a-b\ln r_0)(X-X_g)\right]
\label{rho_02}
\end{equation}
Using equation (\ref {rho_0}) and calling $\rho_0(r_0,X_g)$=$\rho_0(r_0)$
(assuming cylindrical symmetry with respect to the shower axis), we can
write
\begin{equation}
\rho_g(r,\varphi)=\cos\Theta\rho_0(r_0) \exp\left[-(a-b\ln
r_0)(X-X_g)\right]
\label{rho_g}
\end{equation}
The dependence of the right hand side of this equation on $r$ and
$\varphi$ derives from equations (\ref {ro}), (\ref {X}) and (\ref
{zQ}).

In our analysis, the free parameters $a$ and $b$ of equation (\ref {rho_g})
were always evaluated performing least squares fits to data coming from
simulations with the GF disabled (this ensures the cylindrical symmetry of
the shower front plane distributions).
The associated simulations performed in similar conditions, but enabling
the  GF were processed using the same $a$ and $b$ obtained from the ``no
GF" case. This procedure ensures then that the average effect of geometry
and attenuation is properly subtracted, and that the lack of cylindrical
symmetry observed in the corresponding cases is direct consequence of the
GF
deflections. 

\def\journal#1#2#3#4{{\em #1,\/} {\bf #2}, #3 (#4)}

\clearpage
\def\epsfig#1{\epsfbox{#1}}
\begin{figure}
\begin{displaymath}
\hbox{\epsfxsize=10cm\epsfig{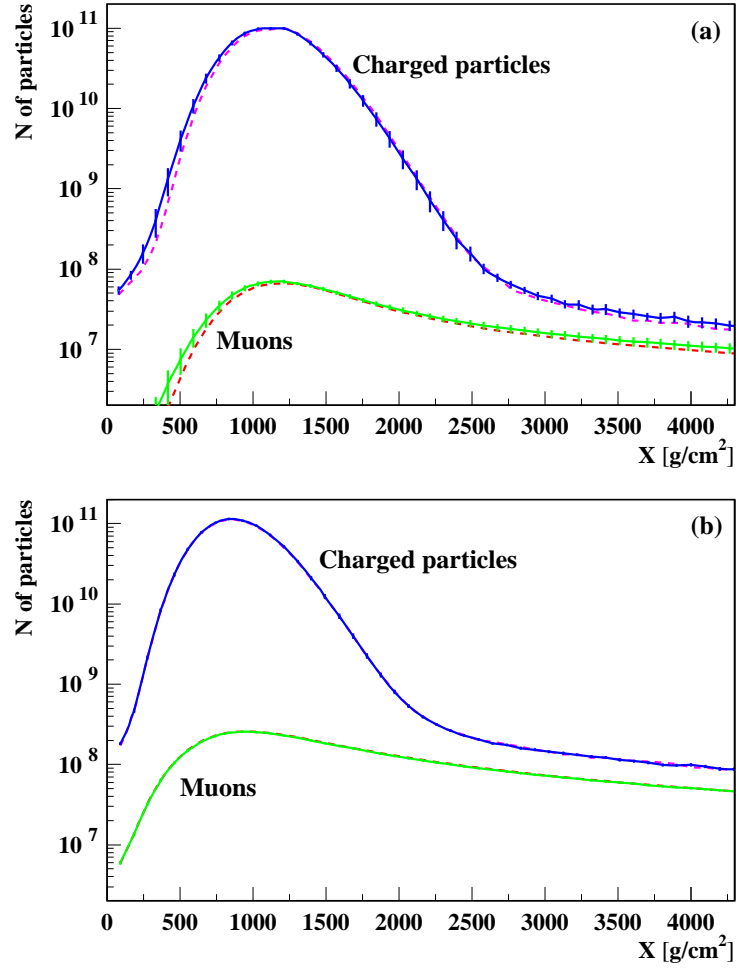}}
\end{displaymath}
\caption{ Longitudinal development of all charged particles and muons,
for $10^{20}$ eV gamma (a) and proton (b) showers. The solid (dashed) 
lines correspond to the case of GF enabled (disabled). The abscissas
represent the slant path along the shower axis, measured from the
injection point. The errors bars correspond to two times the RMS error of
the mean.}
\label{fig:longigp81}
\end{figure}
\begin{figure}
\begin{displaymath}
\hbox{\epsfig{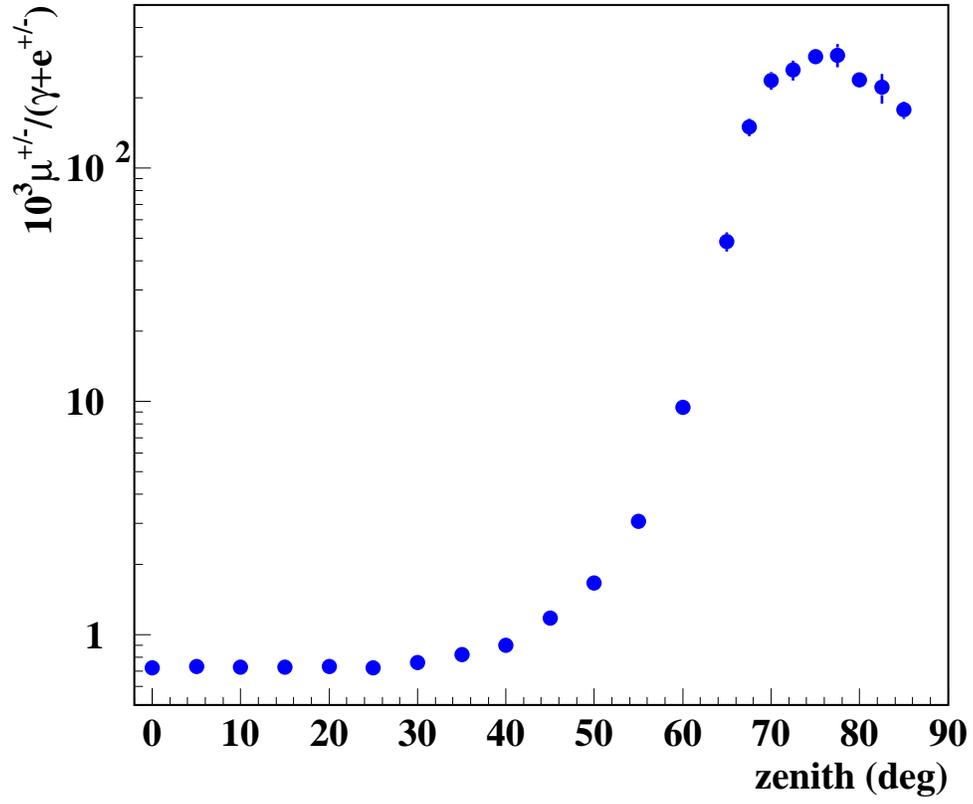}}
\end{displaymath}
\caption{The ratio $\eta$ versus the zenith angle. The simulations
correspond to $3\times10^{20}$ proton showers. Site: El Nihuil.}
\label{fig:ratiovszenithe3}
\end{figure}
\begin{figure}
\begin{displaymath}
\hbox{\epsfig{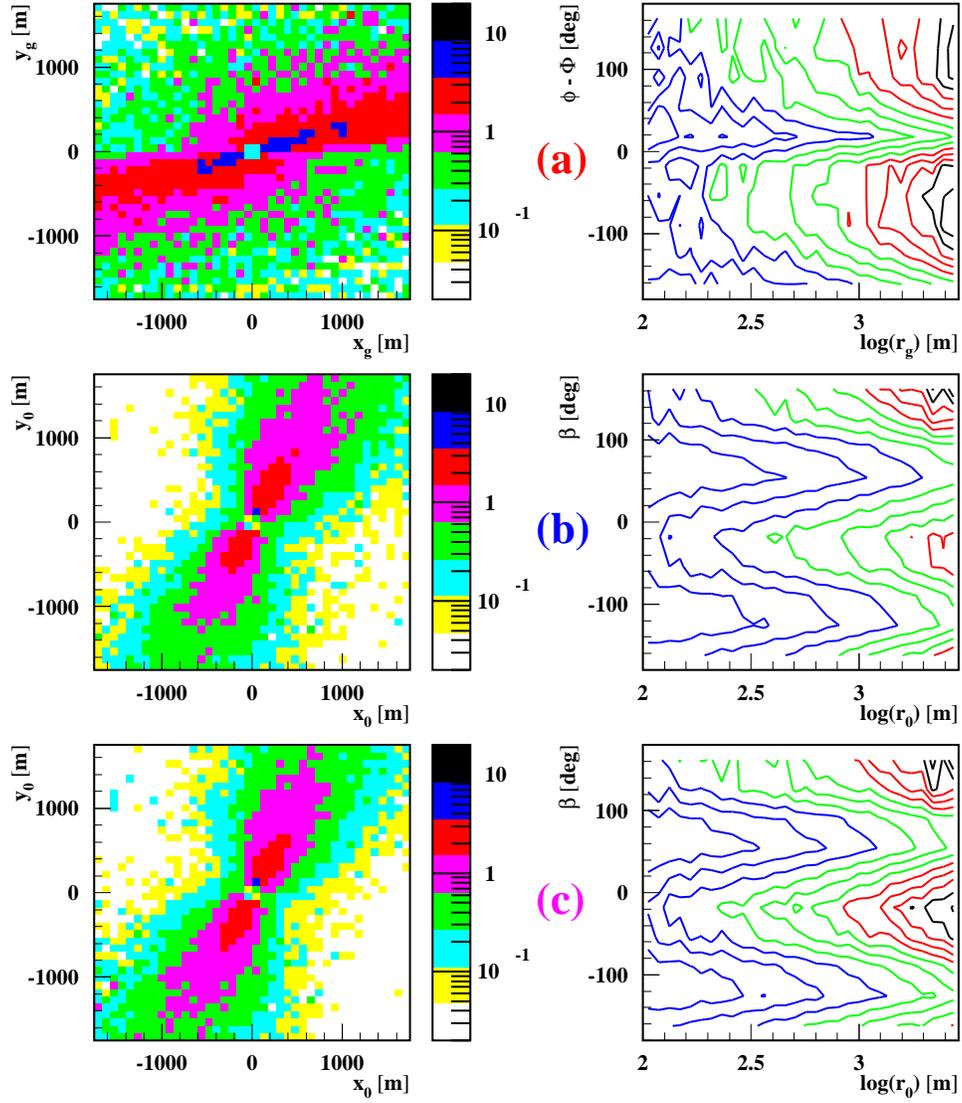}}
\end{displaymath}
\caption{Two dimension lateral distribution of muons for  $10^{20}$ eV
gamma showers, zenith angle $81.5^{\circ}$ at  Millard site.
Injection at the top of the atmosphere. The distributions are represented
as false color diagrams (left column) and contour plots (right column).
(a) Row distribution at ground. (b) Geometrical projection onto the shower
plane (equation (\ref {rho_0})). (c) Geometrical projection plus
attenuation correction (equation (\ref {rho_g})).}  
\label{fig:g100z81conb_mu012}
\end{figure}
\begin{figure}   
\begin{displaymath}
\hbox{\epsfig{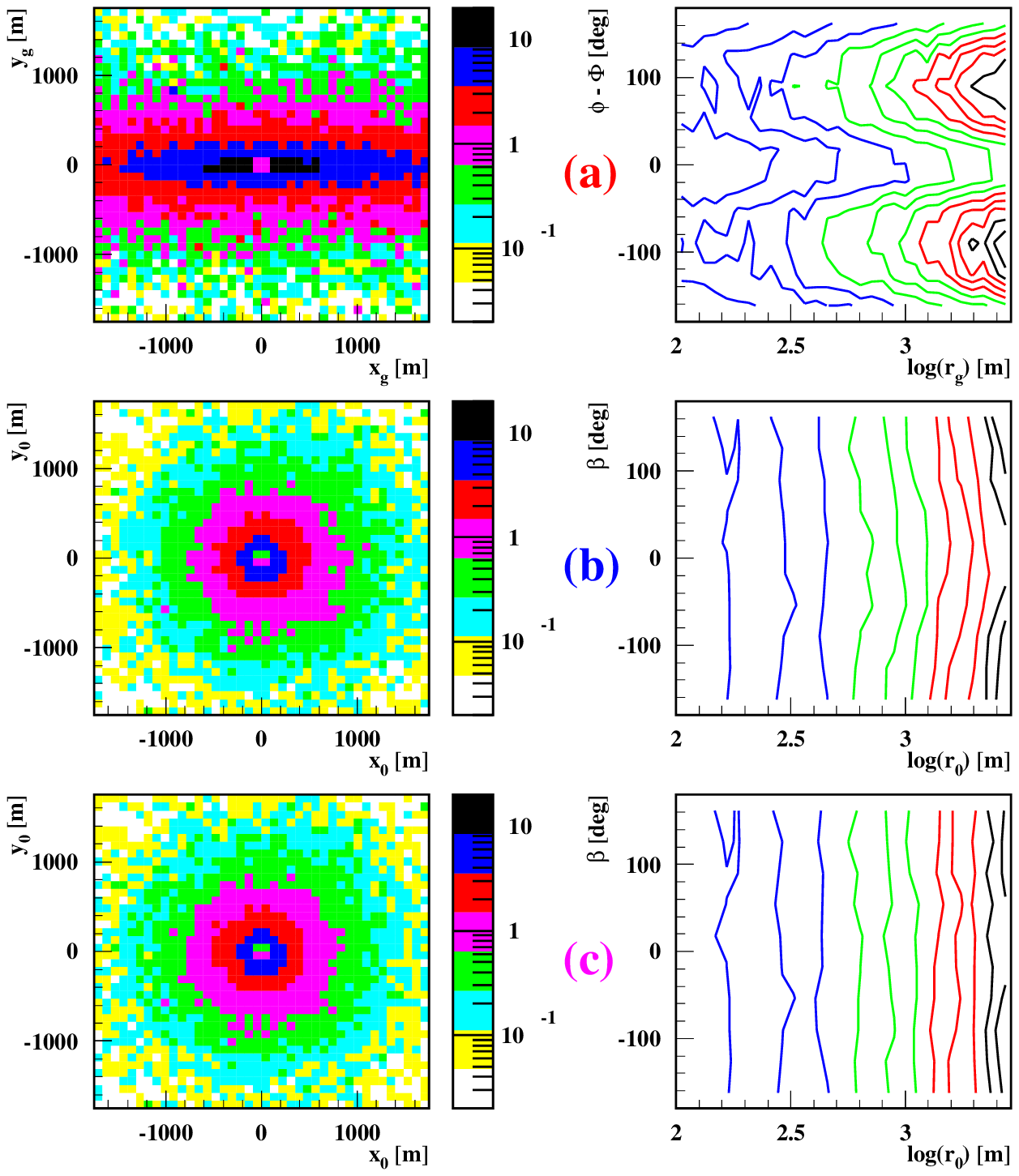}}
\end{displaymath}
\caption{Same as figure \ref {fig:g100z81conb_mu012}, but with the GF
disabled.}
\label{fig:g100z81sinb_mu012}
\end{figure}
\begin{figure}
\begin{displaymath}
\hbox{\epsfig{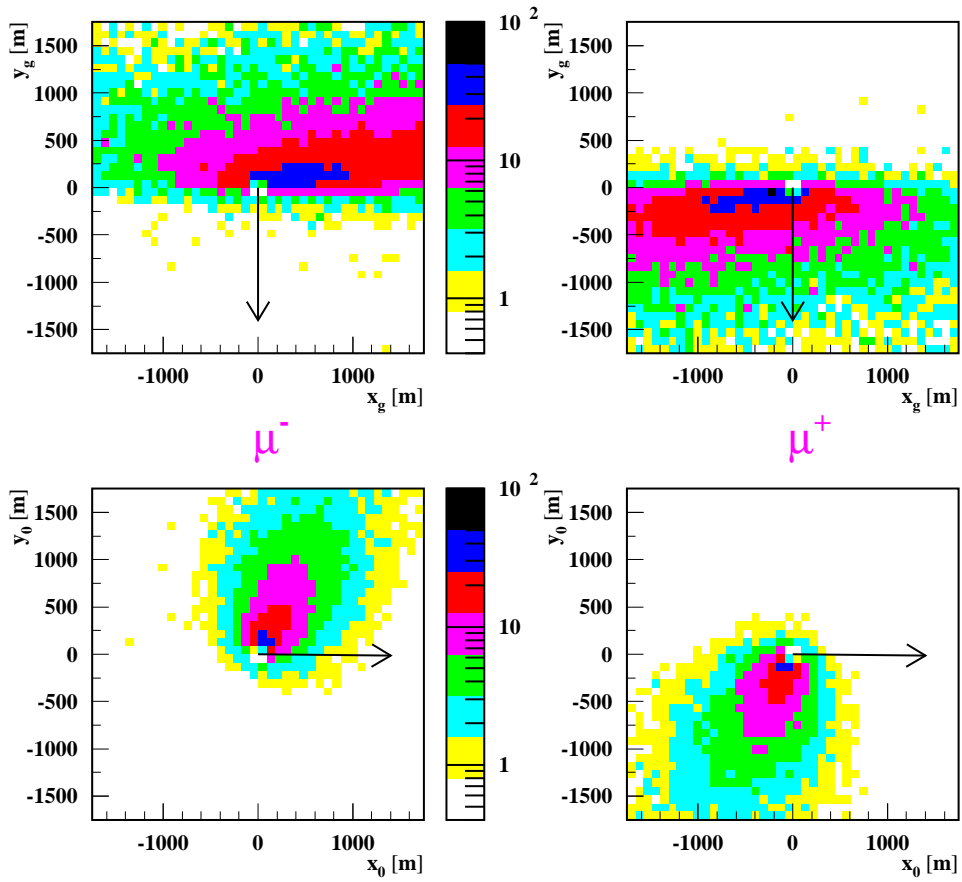}}
\end{displaymath}
\caption{False color representations of the two dimension $\mu^{+}$ and
$\mu^{-}$ density distribution. The upper plots correspond to the row
ground plane distributions, while the lower ones to the corrected shower
front plane distributions (equation (\ref {rho_g})). The arrows indicate
the direction of the H component of the GF (upper plots) or the direction
of the projection of
the GF onto the shower front plane (lower plots). The conditions of the
simulations are the same as in figure \ref {fig:g100z81conb_mu012}, but for 
proton showers and injection altitude 200 g/cm$^2$.}
\label{fig:p100z81conb_mpm02}
\end{figure}
\begin{figure}
\begin{displaymath}
\hbox{\epsfig{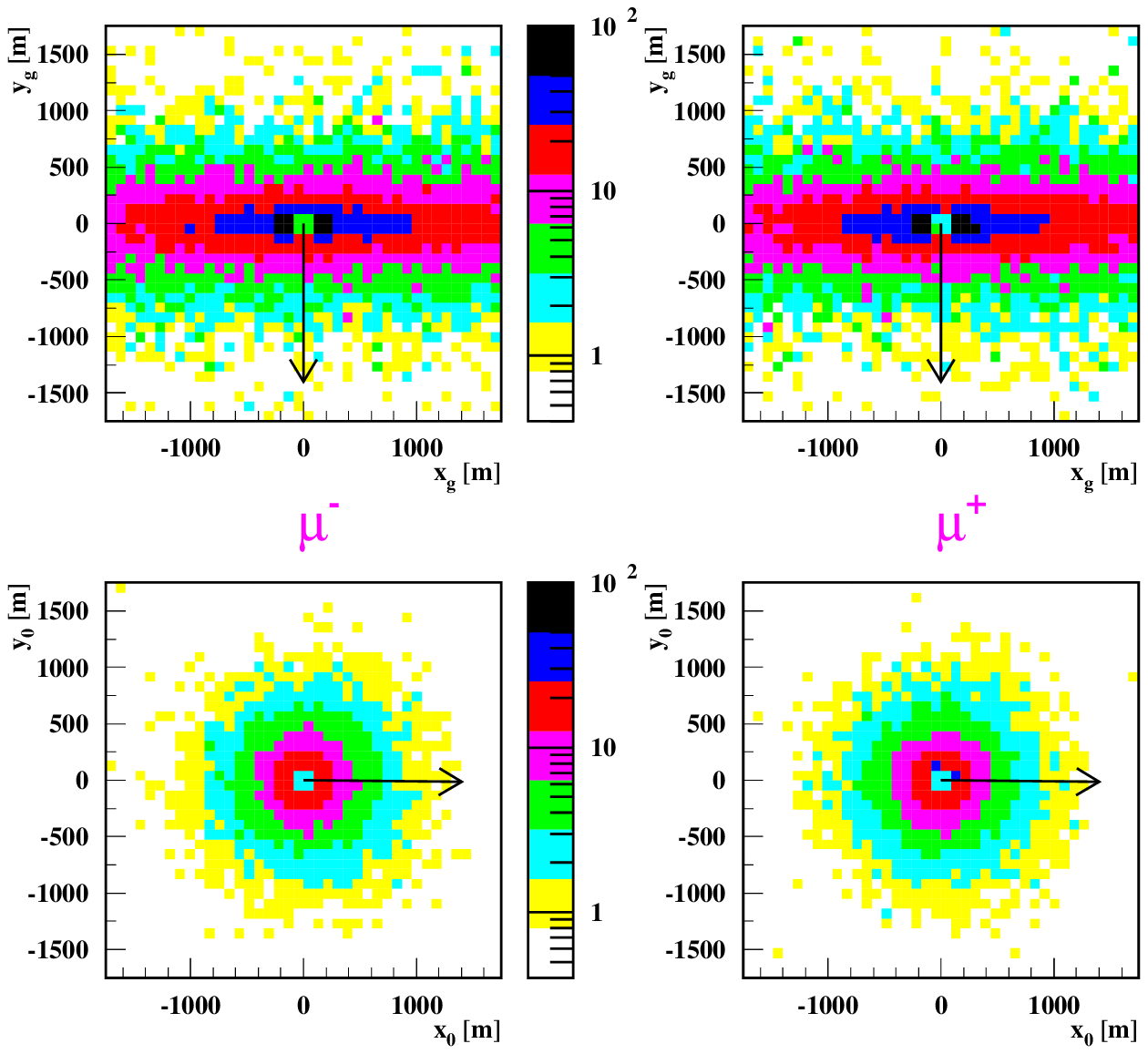}}
\end{displaymath}
\caption{Same as figure \ref {fig:p100z81conb_mpm02}, but with the GF
disabled.}
\label{fig:p100z81sinb_mpm02}
\end{figure}
\begin{figure}
\begin{displaymath}
\hbox{\epsfig{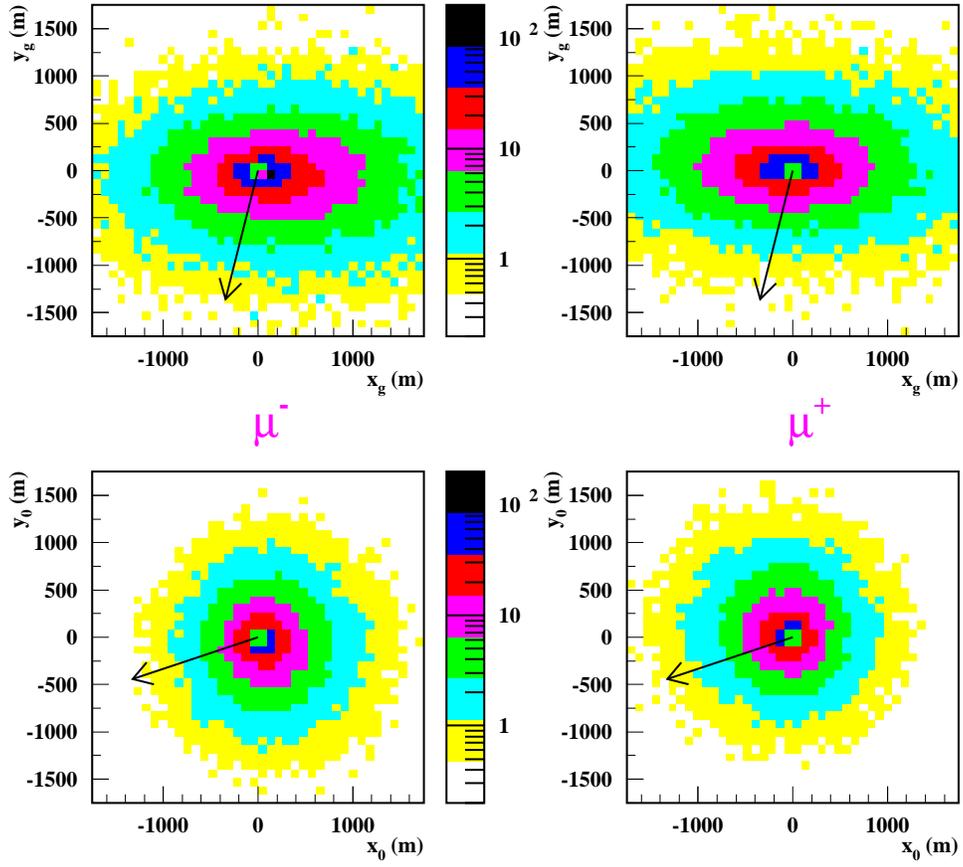}}
\end{displaymath}
\caption{Same as figure \ref {fig:p100z81conb_mpm02} but for
$3\times10^{19}$ eV proton showers, $60^{\circ}$ zenith angle, site El
Nihuil (GF enabled).}
\label{fig:p60n8em_mpm02}
\end{figure}
\begin{figure}
\begin{displaymath}
\begin{array}{cc}
\hbox{\epsfxsize=8cm\epsfig{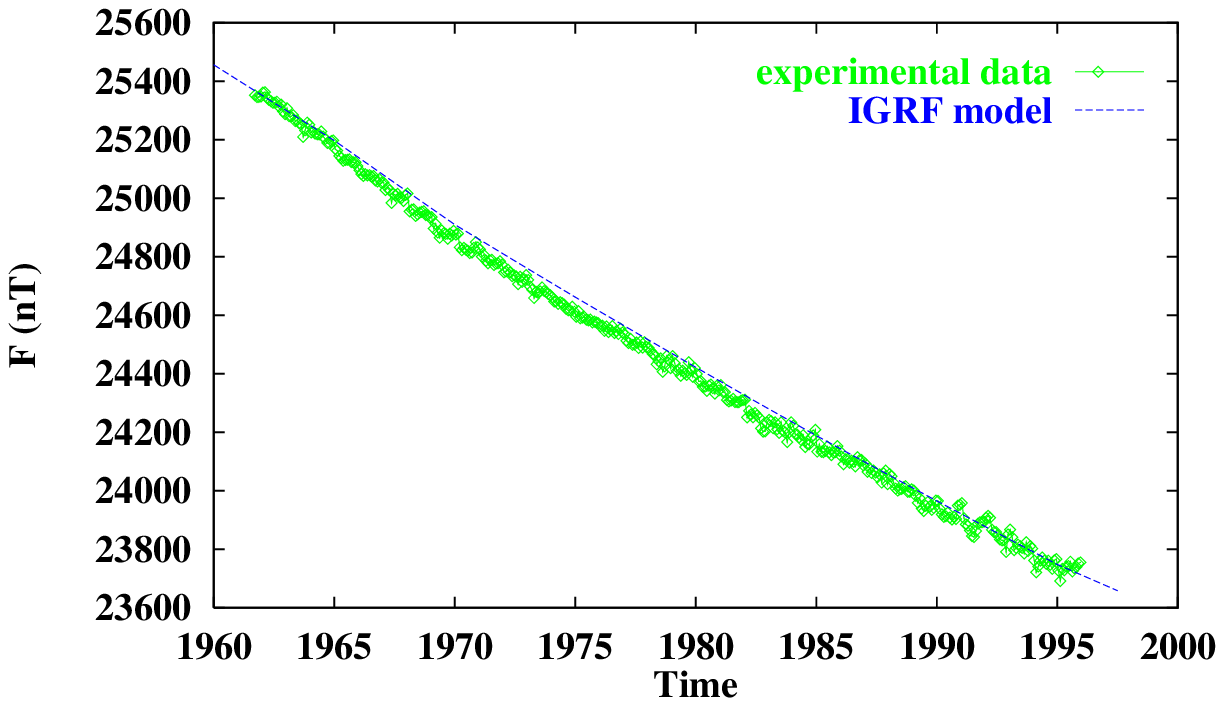}} \\
\hbox{\epsfxsize=8cm\epsfig{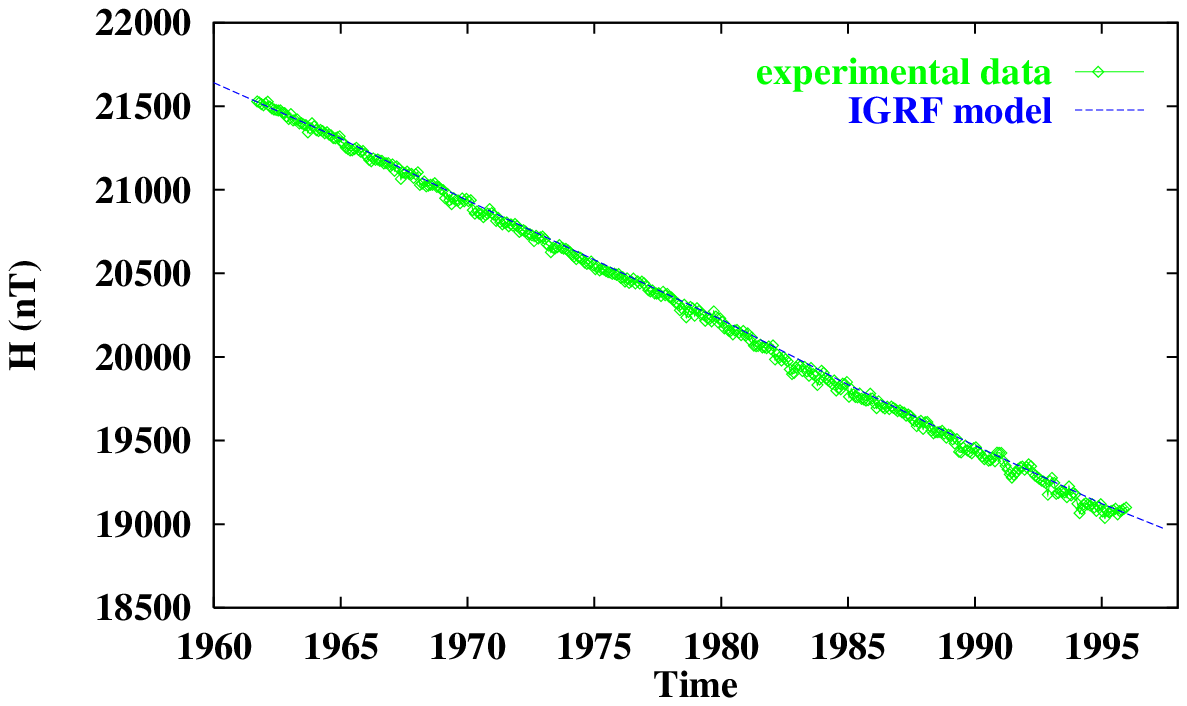}} \\
\hbox{\epsfxsize=8cm\epsfig{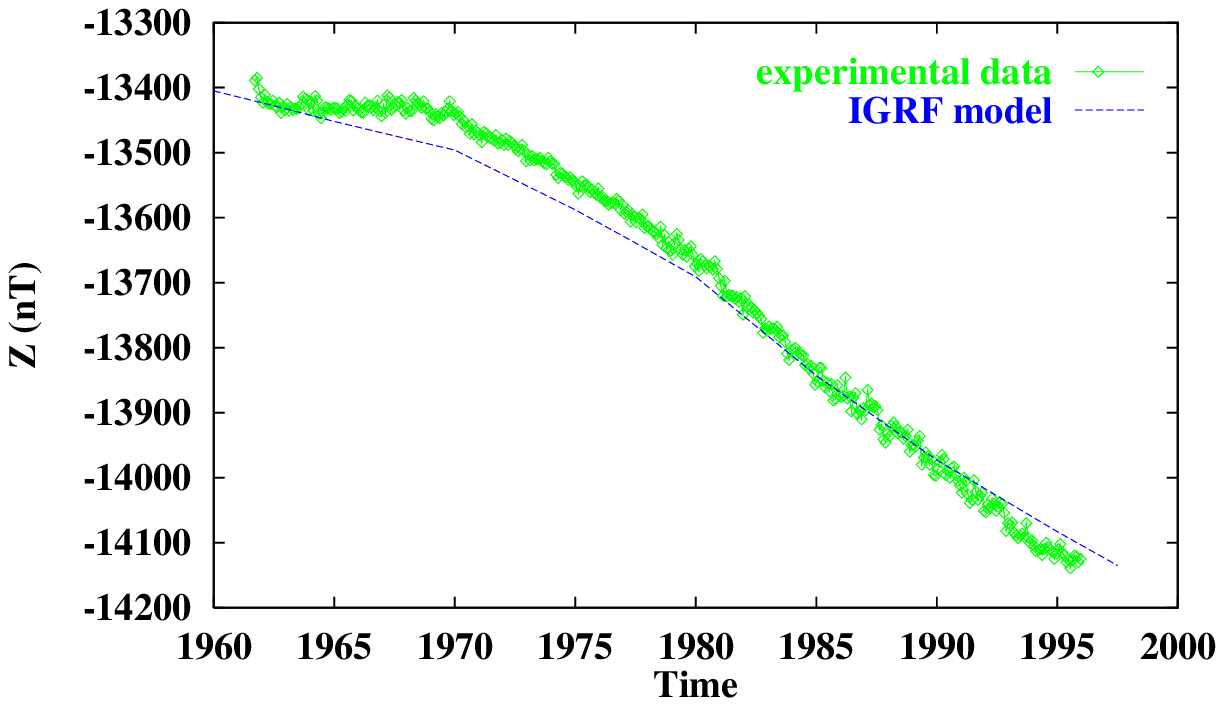}} 
\end{array}   
\end{displaymath}
\caption{Comparison between experimental data and the IGRF model. The
absolute difference between the experimental and the IGRF prediction is
always less than 200 nT.(Experimental data: Las Acacias observatory,
Argentina).}
\label{fig:acacia}
\end{figure}
\begin{figure}   
\begin{displaymath}
\hbox{\epsfig{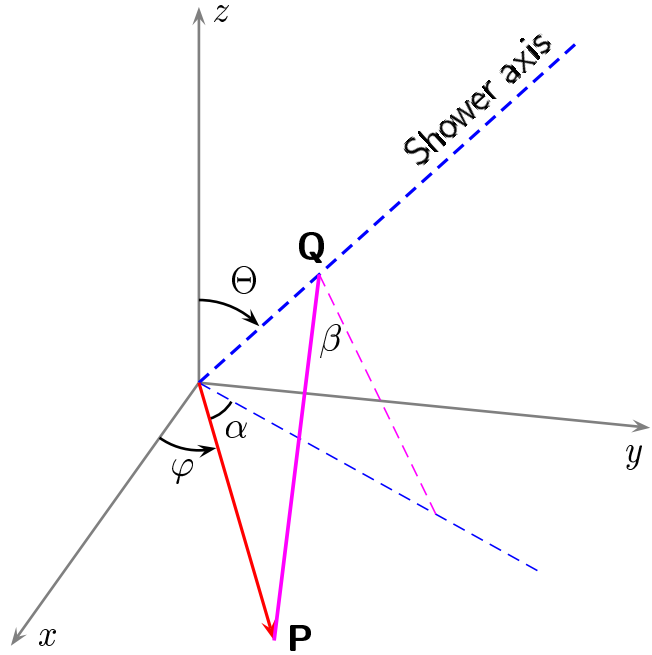}}
\end{displaymath}
\caption{Schematic representation of the polar coordinates
of the point P in the ground  ($r$, $\varphi$) and shower
front ($r_0$, $\beta$) planes. $\alpha=\varphi-\Phi$, where
$\Phi$ is the shower azimuth angle.}
\label{fig:geometry}
\end{figure}
\end{document}